\documentclass{article}
\usepackage[subpreambles=true]{standalone}
\usepackage[utf8]{inputenc}
\usepackage{feynmf}
\usepackage{amsmath,amssymb}
\usepackage{bbold}
\usepackage{graphicx} 
\usepackage{epstopdf}
\usepackage{subcaption}
\usepackage{mathtools}
\usepackage{mathrsfs}
\usepackage[margin=1.0in]{geometry}
\usepackage{fancyhdr}
\usepackage{lmodern}
\usepackage{subfiles}
\usepackage[absolute,overlay]{textpos}
\usepackage{anyfontsize}
\usepackage[colorlinks]{hyperref}
\usepackage{adjustbox}
\usepackage{authblk}
\hypersetup{linkcolor=blue}
\allowdisplaybreaks

\newcommand{\A}{\mathcal{A}}
\newcommand{\B}{\mathcal{B}}
\newcommand{\C}{\mathcal{C}}
\newcommand{\D}{\mathcal{D}}
\newcommand{\E}{\mathcal{E}}
\newcommand{\M}{\mathcal{M}}
\newcommand{\f}[1]{f^{#1}}

\newcommand{\pole}{\Omega}
\newlength\longest

\title{\textbf{The unitarity of a spontaneously broken $SU(2)$ theory using unitary-gauge diagrams}}
\author{Jochem Kip\thanks{Jochem.kip@ru.nl} \hspace{1pt} and Ronald Kleiss\thanks{R.Kleiss@science.ru.nl}}
\affil{Institute for Mathematics, Astrophysics and Particle Physics, Radboud
University Nijmegen, Heyendaalseweg 135, Nijmegen, the Netherlands
}
\date{}
\begin{document}

\maketitle
\begin{abstract}
A spontaneously broken SU(2) theory is the simplest generalization of the Abelian Higgs model, containing three equally massive vector bosons and a single Higgs scalar. A strictly diagrammatic proof is presented of the tree-level unitarity of this model in the unitary gauge, i.e. employing only physical fields. A new Ward-like identity is used to show that the high-energy behaviour of all amplitudes cannot be more than quadratic; the use of generating functions for all off-shell amplitudes then leads to the unitarity proof.
\end{abstract}
\newpage

\section{Introduction}
In this paper the unitarity of a spontaneously broken $SU(2)$ theory is investigated at tree level using the unitary gauge, solely using Feynman diagrams. It is well-known that the standard model and number of embedded theories are internally consistent; QCD or QED as examples. The unitarity of the Abelian Higgs model, containing solely the Higgz and $Z$ bosons, has been proven\cite{abelianhiggs}. In this paper an extension towards an $SU(2)$ theory is made; the simplest of the special unitary groups, containing the $W^+$, $W^-$, $Z$, and $H$ bosons and their interactions. Naturally the $W^\pm$ bosons are not electrically charged, seeing as there is no photon, but this notation will nevertheless be kept to refer to the assigned charge of the $SU(2)$ group, e.g. isospin. Especially important is the energy behaviour of any on-shell amplitude. The unitarity of a spontaneously broken $SU(2)$ theory is, while not trivial, well understood\cite{THOOFT1972189}. The conventional way of approaching such problems is by starting with an unbroken Lagrangian in some conveniently chosen gauge and subsequently breaking it via the Higgs mechanism. Alternatively, it is possible to demand that only physical fields appear, i.e. using the unitary gauge, and working solely via the resulting Feynman diagrams. The unitary gauge is conventionally regarded as 'non-renormalizable'; this is, of course, wrong\footnote{Only a single quartic counterterm is needed for full renormalizability\cite{Grosse-knetter}}. Within this paper it will be shown that, at least at tree level, all dangerous terms resulting from the unitary gauge vanish .\\
Unitarity dictates that the cross section of a scattering amplitude ($2\rightarrow n-2$) has to scale as $E^{-2}$ at high energies when all angles are fixed\cite{Peskin:1995ev,Itzykson:1980rh}. Moreover, power counting shows that the phase space scales as $E^{n-4}$, thus the upper-bound for an $n$-particle amplitude must be:
\begin{align}
    \M_n \sim E^{4-n} \label{eq:unitarity_bound}
\end{align}
This bound is not trivially obtained; the non-vanishing vector boson propagators, the longitudinal propagators present in massive gauge theories, and the Yang-Mills three-point interaction, which contains momentum terms, all provide increments to the energy dependence of an amplitude. For example, the number of tree-level diagrams for the process $W^+ W^- \rightarrow 2W^+2W^-2Z4H$ is over\footnote{4,068,097,116 to be precise.} $4\hspace{0.5em}10^9$, of which the highest-scaling diagrams have an a priori $E^{14}$ dependence, as dictated by power counting. More generally, for an amplitude involving $s$ $W^+W^-$ pairs, $i$ $Z$'s, and $j$ $H$ bosons, the highest-scaling diagrams have a leading contribution of $E^{4s+2i-2}$, thus requiring the energy dependence to be dropped by a factor of $E^{4s+3i+j-6}$. Naturally, systematic cancellations are needed, especially seeing as manual computation of diagrams becomes fairly cumbersome if the number of particles in an amplitude increases. Fortunately, it is not necessary to prove all cancellations down to the bound of eqn \eqref{eq:unitarity_bound}; cancellations up to $E^{-1}$ are sufficient for any amplitude with five or more external particles\footnote{Four-particle amplitudes are of course already safe at $E^0$.}\cite{abelianhiggs}.\\
This paper is structured as follows. First the Feynman rules will be established, in addition to some conventions regarding notation. Next, a number of general cancellation mechanisms will be laid out. With which it is shown, in a manner reminiscent of the Ward-Takahashi identity, that the energy behaviour of an arbitrary off-shell amplitude is $E^2$ at the highest. The $E^2$ terms are then shown to vanish via the solutions of the Swinger-Dyson equations for off-shell amplitudes at leading order. These solutions are then subsequently used to also cancel the remaining $E^0$ terms by taking subleading contributions to the off-shell amplitudes. Finally, the deformation proof of \cite{abelianhiggs} is shortly given to complete the proof that a spontaneously broken $SU(2)$ theory is unitary at tree level.

\section{Feynman rules for a spontaneously broken $SU(2)$ theory}
The propagator of the $W^+$, $W^-$, and $Z$ boson within the unitary gauge is given by:\\
\begin{align}
    \adjustbox{valign=c}{\includegraphics[scale=.65]{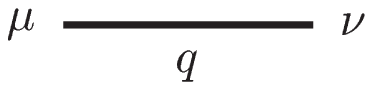}}=\frac{i}{q^2-m^2}\left(-g^{\mu\nu}+\frac{q^\mu q^\nu}{m^2}\right)
\end{align}
The convention is adopted where the propagators of the $W$'s are oriented, while an unoriented line indicates a $Z$. The orientation of the $W$ propagators is an immidiate result of charge conservation\footnote{Note that this is of course not electrical charge, but rather isospin.}. Moreover, the vector bosonic propagator is explicitly split into two terms, unless stated otherwise:
\begin{align}
&\adjustbox{valign=c}{\includegraphics[scale=.7]{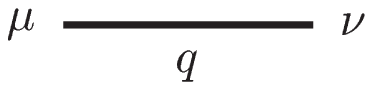}}= -g^{\mu\nu} \frac{i}{q^2-m^2} & \adjustbox{valign=c}{\includegraphics[scale=.7]{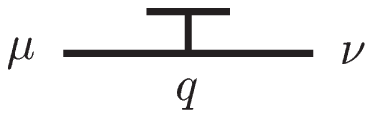}}= \frac{q^\mu q^\nu}{m^2}\frac{i}{q^2-m^2}
\end{align}
The Higgs propagator is given by:
\begin{align}
\adjustbox{valign=c}{\includegraphics[scale=.7]{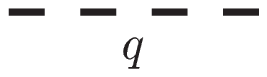}}=\frac{i}{q^2-M^2}
\end{align}
The vector boson and Higgs masses are denoted by $m$ and $M$ respectively. The couplings of a spontaneously broken $SU(2)$ theory can either be obtained from the proper Lagrangian, or by setting the Weinberg angle and electric charge to zero in electroweak theory. Both methods result in the following Feynman rules\cite{Peskin:1995ev}:
\begin{align}
    &\adjustbox{valign=c}{\includegraphics[scale=.4]{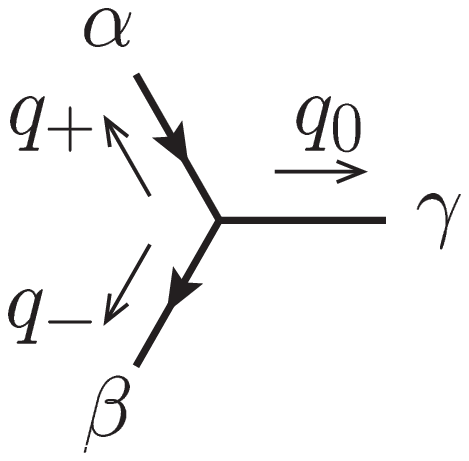}}&
    &\adjustbox{valign=c}{\includegraphics[scale=.4]{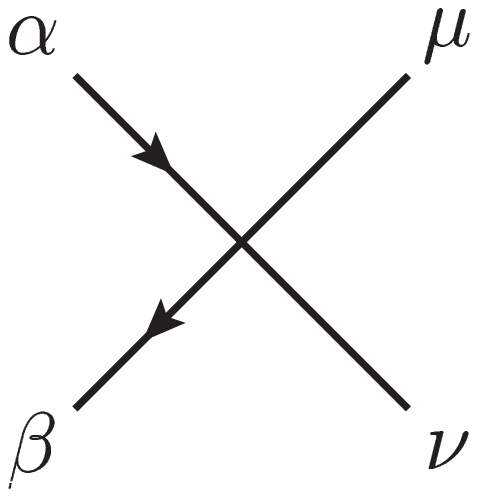}}& 
    &\adjustbox{valign=c}{\includegraphics[scale=.4]{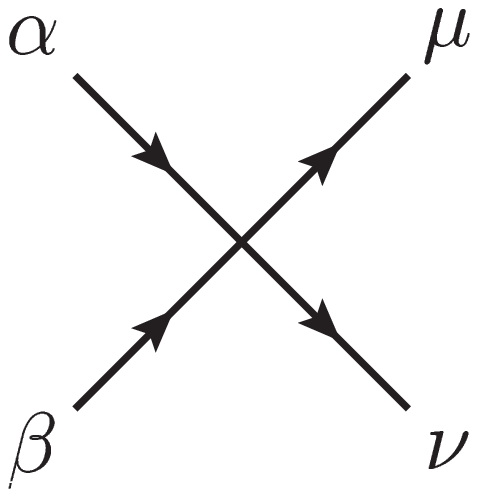}}& \nonumber \\
    &=-2igm Y(q_+,\alpha;q_-,\beta;q_0,\gamma) && =-4ig^2m^2 X^{\alpha\beta\mu\nu} && =4ig^2m^2X^{\alpha\beta\mu\nu} & \nonumber\\
    &\adjustbox{valign=c}{\includegraphics[scale=.4]{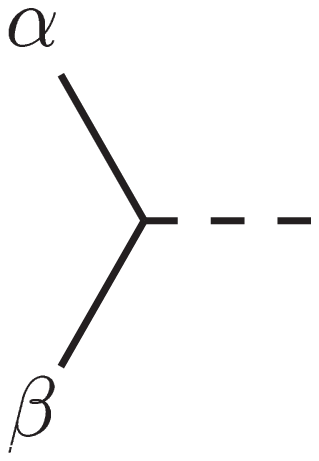}}&
    &\adjustbox{valign=c}{\includegraphics[scale=.4]{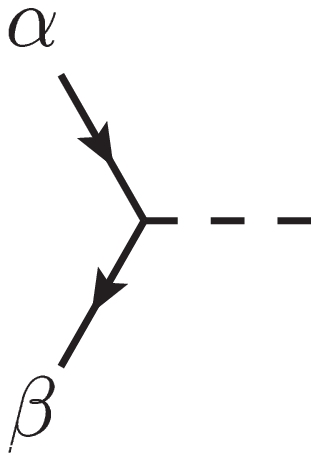}}&
    &\adjustbox{valign=c}{\includegraphics[scale=.4]{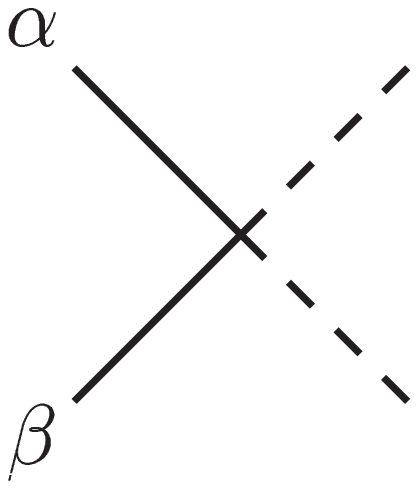}}& \nonumber\\
    & =2igm^2g^{\alpha \beta} && =2igm^2g^{\alpha \beta} && =2ig^2m^2g^{\alpha \beta} & \nonumber \\
    &\adjustbox{valign=c}{\includegraphics[scale=.4]{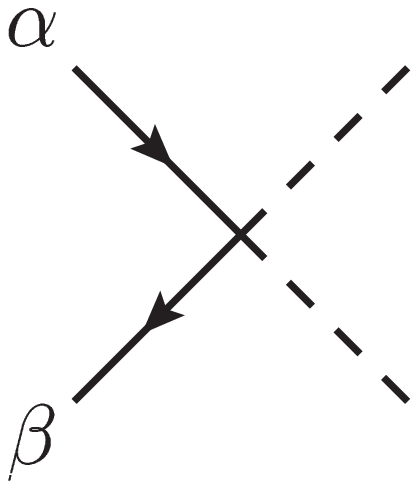}} &
    &\adjustbox{valign=c}{\includegraphics[scale=.4]{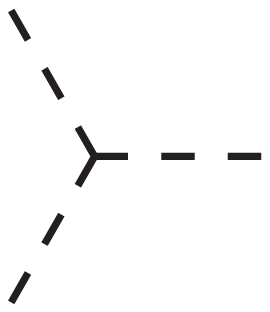}}&
    &\adjustbox{valign=c}{\includegraphics[scale=.4]{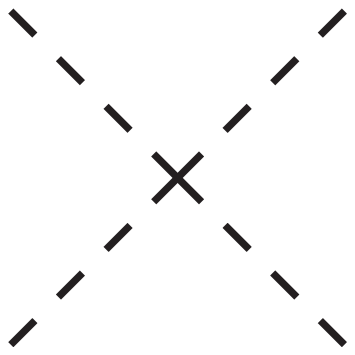}}& \nonumber\\
    & =2igm^2g^{\alpha \beta} && = -3igM^2 && = -3ig^2M^2 &\label{eq:Feynman_rules}
\end{align}
The dimensionful coupling constant is given by $g^2=\sqrt{2}\hspace{0.2em}G_F$, and the $Y(q_+,\alpha;q_-,\beta;q_0,\gamma)$ and $X^{\alpha \beta \mu \nu}$ functions are  defined as:
\begin{align}
    &Y(q_+,\alpha;q_-,\beta;q_0,\gamma)=(q_+-q_-)^\gamma g^{\alpha \beta}+(q_--q_0)^\alpha g^{\beta \gamma}+(q_0-q_+)^\beta g^{\alpha \gamma}\\
    &X^{\alpha \beta \mu \nu}=2g^{\alpha \beta}g^{\mu\nu}-g^{\alpha \mu}g^{\beta \nu}-g^{\alpha \nu}g^{\beta\mu}
\end{align}
The coupling containing the $Y$-function is the Yang-Mills three-point vertex (TPV), while couplings with the $X$-function are Yang-Mills four-point vertices (FPV). Furthermore, unless otherwise specified, all external particles are longitudinally polarized, in which the following convention is used:
\begin{align}
    \epsilon_L^\mu(q,t')=\frac{1}{m}\left(q^\mu-\frac{m^2}{q\cdot t'}t'^\mu\right)\equiv \frac{1}{m}\left(q^\mu-t^\mu\right) \label{eq:Longitudinal_polarization_definition}
\end{align}
Where $t'^\mu$ is a light-like vector that can be freely chosen\footnote{All $t'^\mu$ vectors are chosen to be the same for simplicity.} . \\
Finally, while not a Feynman rule, whenever a (sub)amplitude is contracted with its complete momentum, this is referred to as a handlebar operation and diagrammatically represented as:
\begin{align}
     &\adjustbox{valign=c}{\includegraphics[scale=.7]{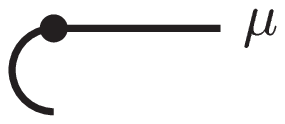}}=q^\mu
\end{align}

\section{$E^2$ as an upper bound on the energy scaling of an amplitude}
The $Y$-function has three different terms, all uniquely specified by the metrics. Diagrammatically the terms will be differentiated by a perpendicular line through the vertex leg whose Lorentz-index is not contained in the metric. By performing the handlebar operation upon all terms and rewriting the resulting expression in a convenient manner, they read\footnote{In the case of mixing, e.g. $SU(2)\times U(1)$, this is of course no longer true}:
\begin{align}
    &\adjustbox{valign=c}{\includegraphics[scale=.4]{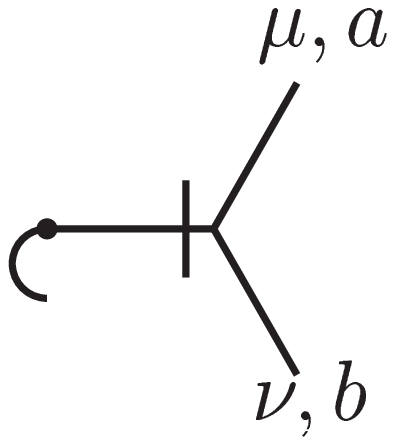}}= -ig'\f{abi}g^{\mu \nu}((q_a^2-m^2)-(q_b^2-m^2)),& &\adjustbox{valign=c}{\includegraphics[scale=.4]{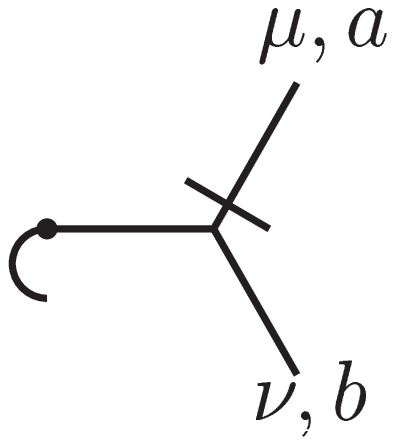}}+ \adjustbox{valign=c}{\includegraphics[scale=.4]{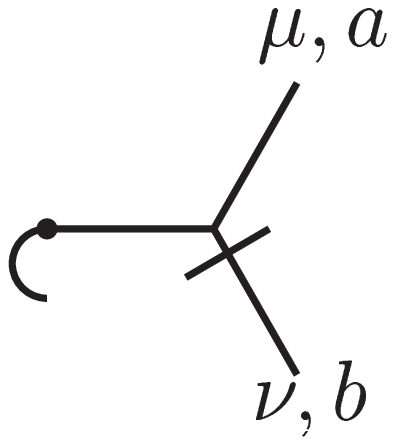}}=-ig'\f{abi}(q_b^\mu q_b^\nu -q_a^\mu q_a^\nu) \label{eq:TPV_split}
\end{align}
The first expression can be seen to contain the denominator of the propagator of particles $a$ and $b$, while the second expression contains $q_a^\mu$ and $q_b^\nu$; the handlebars of particle $a$ and $b$ respectively. Here $f^{abc}$ is of course the Levi-Civita symbol, which are the structure constants for $SU(2)$. The arrows denoting the $W^+$ and $W^-$ bosons are omitted here such that not all possible diagrams need to be explicitly written out.\\
When two TPV's are directly connected, the denominator of their connecting propagator is canceled when the appropriate TPV term is taken in combination with a handlebar, i.e. the first term of equation \eqref{eq:TPV_split}. When all three possible diagrams and a handlebarred FPV are added together, and only the metric part of the propagator is used, the resulting expression is zero:
\begin{align}
\adjustbox{valign=c}{\includegraphics[scale=.4]{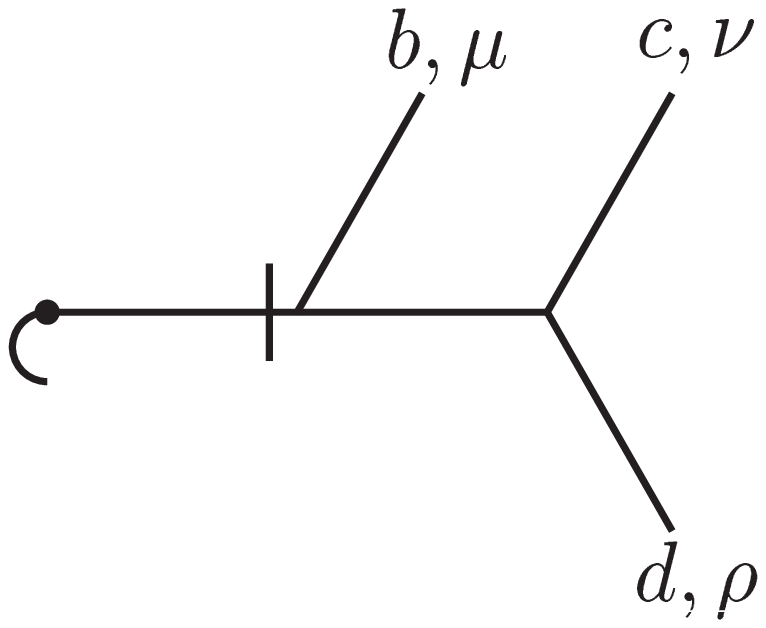}}+
\adjustbox{valign=c}{\includegraphics[scale=.4]{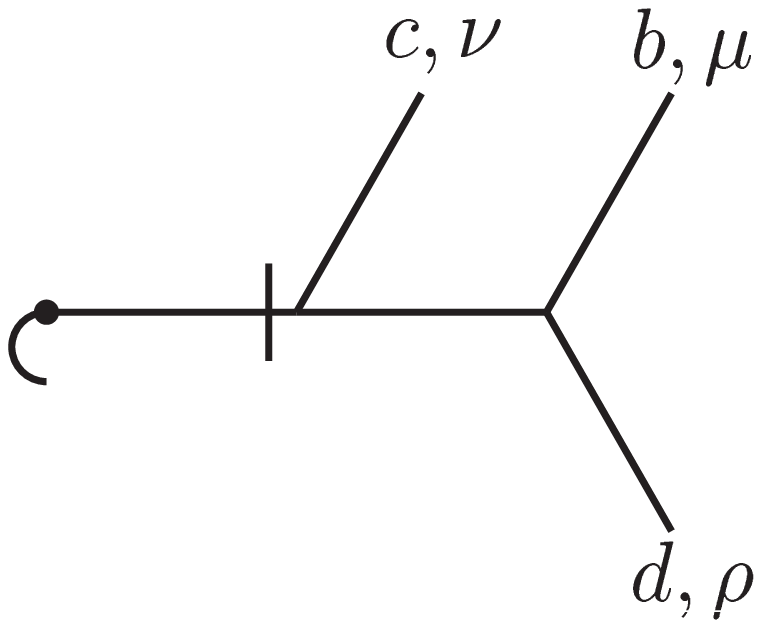}}+
\adjustbox{valign=c}{\includegraphics[scale=.4]{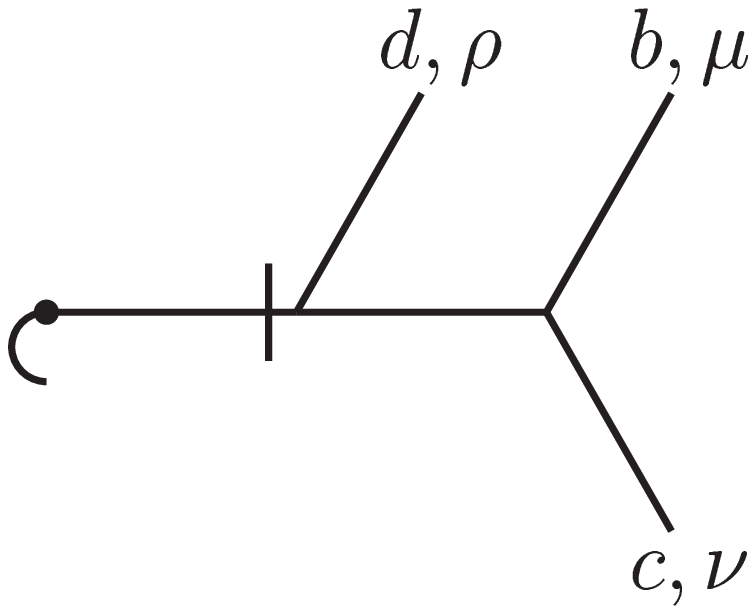}}+
\adjustbox{valign=c}{\includegraphics[scale=.4]{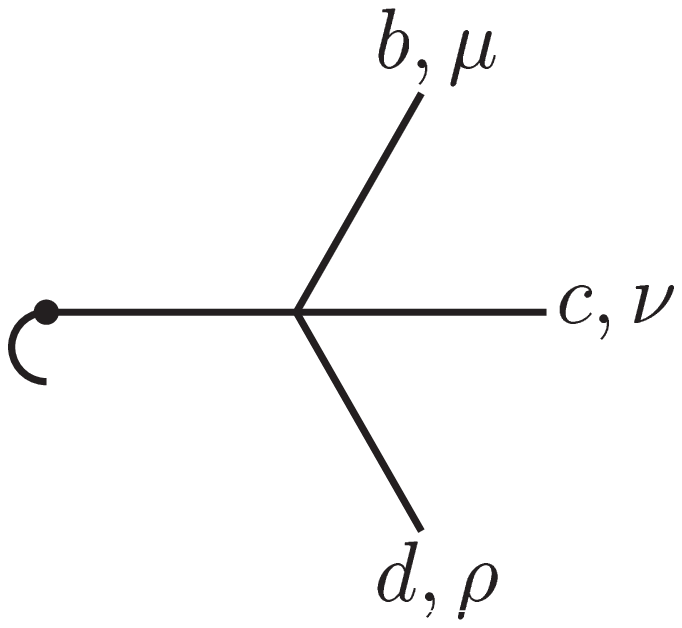}}=0 \label{eq:TPV_TPV_cancel}
\end{align}
Similarly, all four combinations of a TPV directly connected to an FPV can be summed, which, when again only the metric term of the propagator is used, is also zero:
\begin{align}
\adjustbox{valign=c}{\includegraphics[scale=.4]{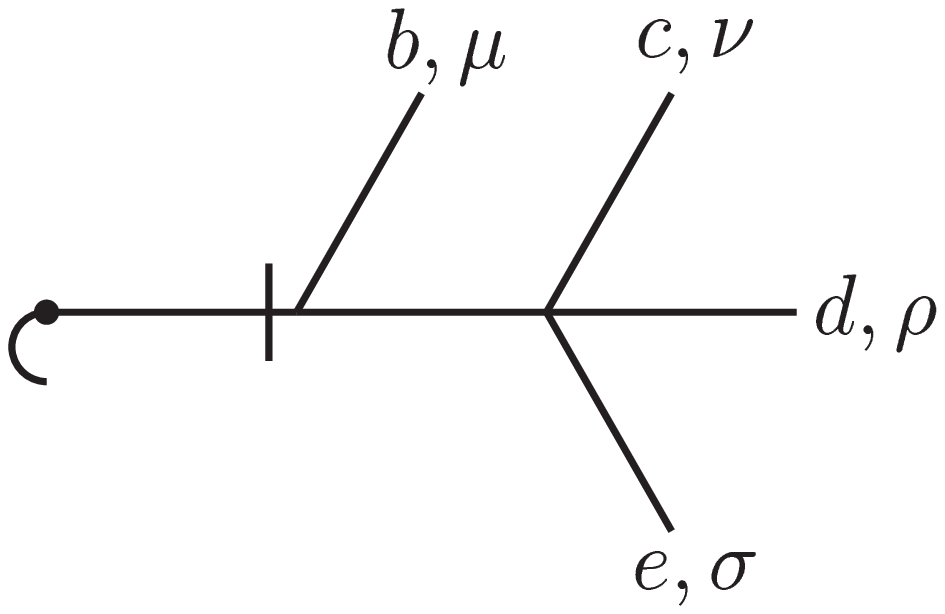}}+
\adjustbox{valign=c}{\includegraphics[scale=.4]{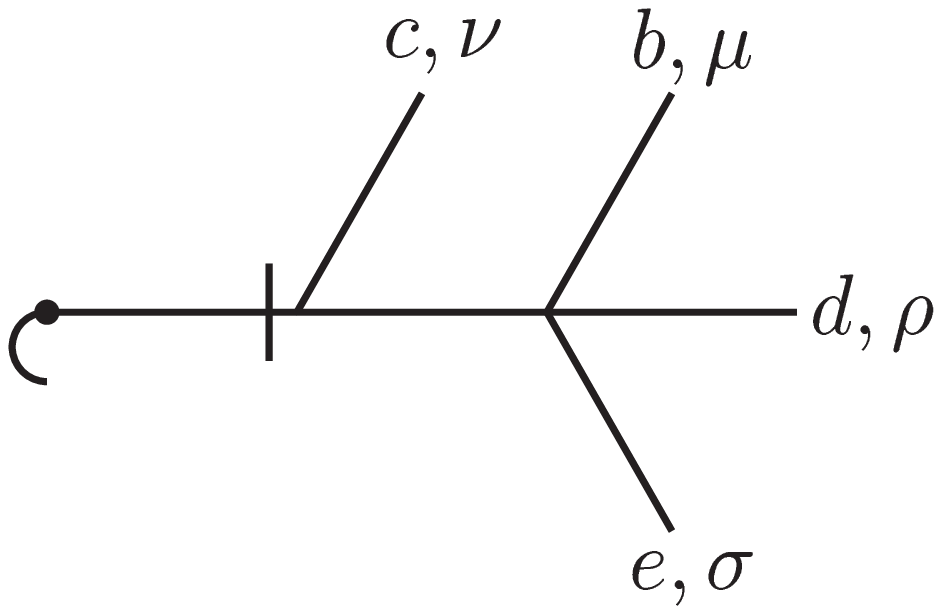}}+\nonumber \\
\adjustbox{valign=c}{\includegraphics[scale=.4]{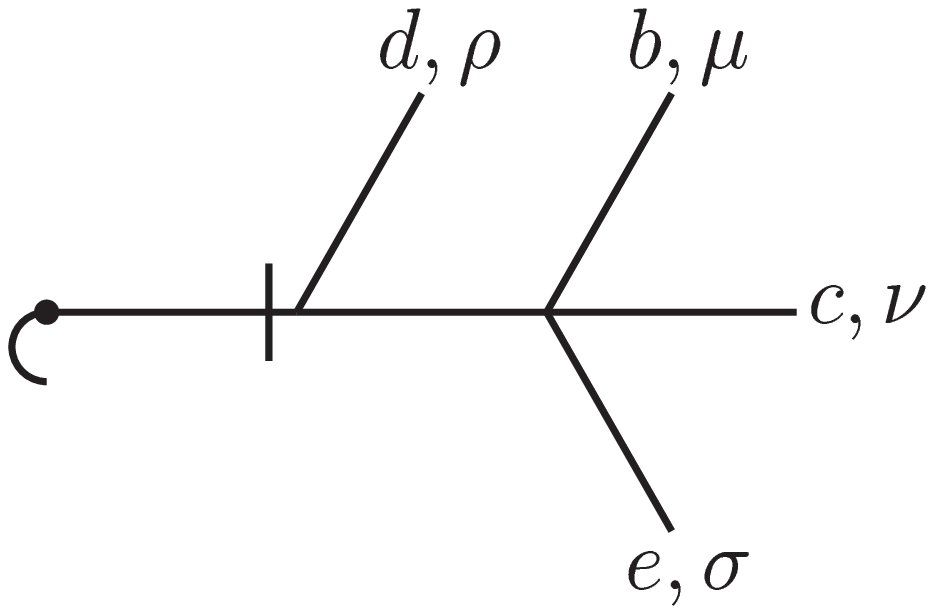}}+
\adjustbox{valign=c}{\includegraphics[scale=.4]{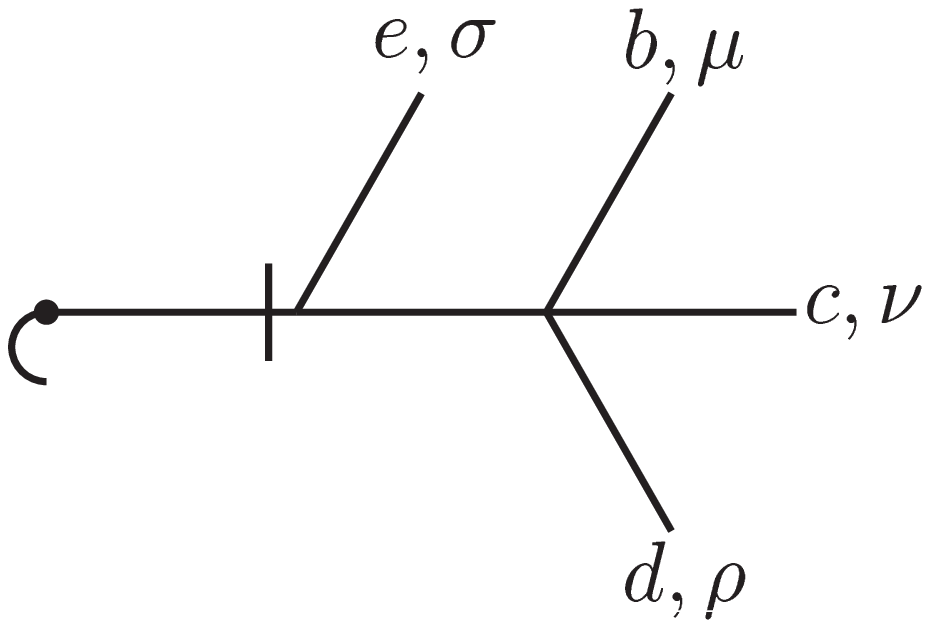}}=0\label{eq:TPV_FPV_cancel}
\end{align}
By adopting a proof reminiscent of a Schwinger-Dyson equation, the aforementioned cancellations and manipulations are sufficient to inductively prove that a handlebar placed upon an amplitude consisting solely of Yang-Mills vertices (YM-amplitude) is zero:
\begin{align}
    \includegraphics[scale=.5]{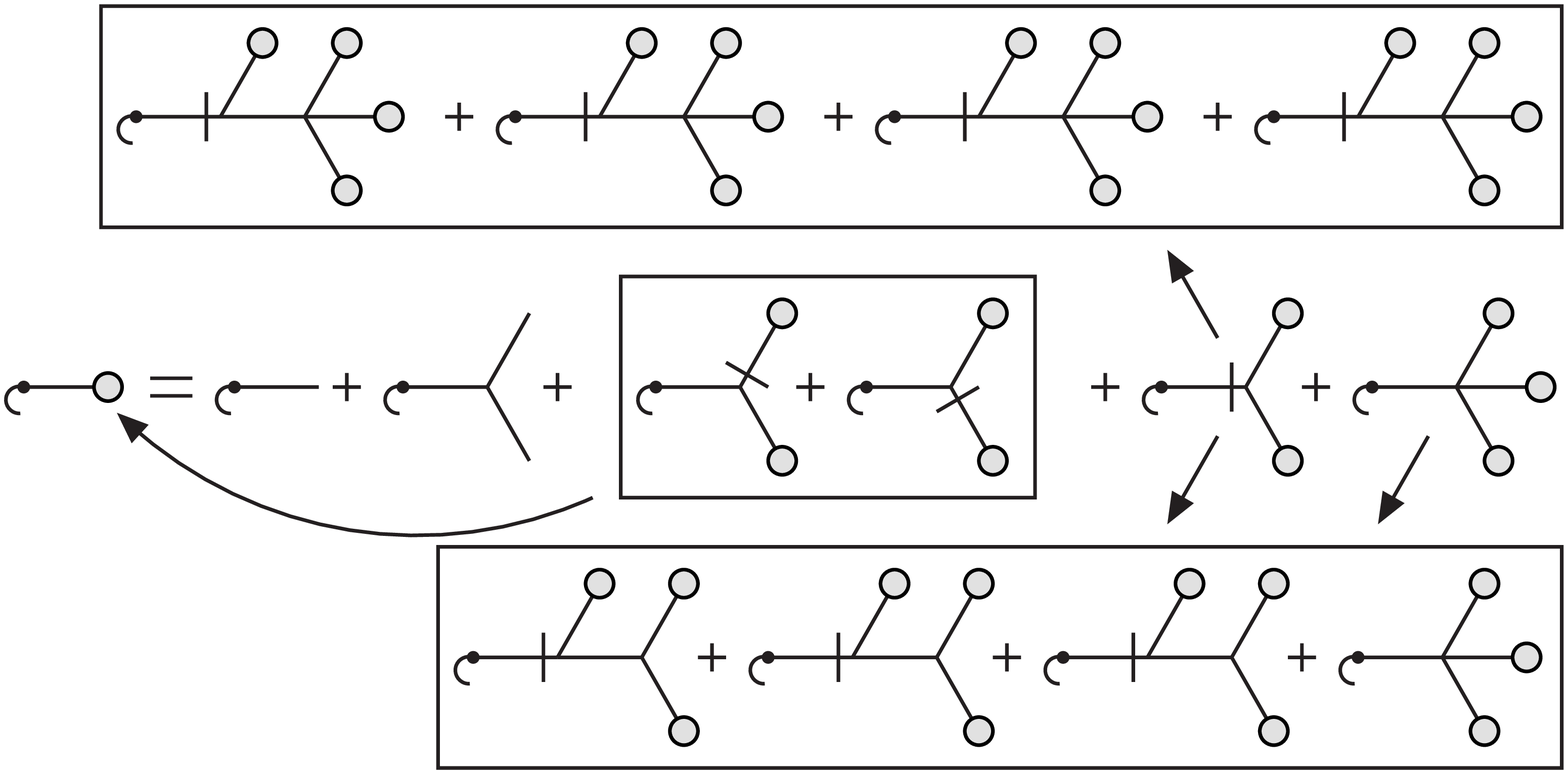}\nonumber
\end{align}
When moving into a YM-amplitude there are four distinct possible first encounters; an external particle, a TPV with two external particles, a TPV in which at least one leg is not an external particle, or an FPV. The first two possibilities are, when handlebarred, trivially zero. When a TPV with one or no external particles is encountered, it is split into the combinations of equation \eqref{eq:TPV_split}; the two terms that yield a handlebar iterate the same process one step further, while the term that annuls the denominator of a connecting propagator is used to cancel a combination of two vertices, either as in equation \eqref{eq:TPV_TPV_cancel}, which includes the FPV, or \eqref{eq:TPV_FPV_cancel}, depending on the ensuing vertex. Therefore, a handlebar on an arbitrary YM-amplitude always gives zero\footnote{Again, only when all vector boson masses are equal.}. \\
When Higgs-vector boson couplings are taken into account it is no longer automatically true that a TPV is directly connected to another Yang-Mills vertex, whereby the cancellations of equations \eqref{eq:TPV_TPV_cancel} and \eqref{eq:TPV_FPV_cancel} are void. However, any such diagrams in which an arbitrary number of Higgs-vector boson couplings are between a TPV and a TPV or FPV are zero by virtue of inter-diagrammatic cancellations. Consider the following two diagrams:
\begin{align}
    &\adjustbox{valign=c}{\includegraphics[scale=.4]{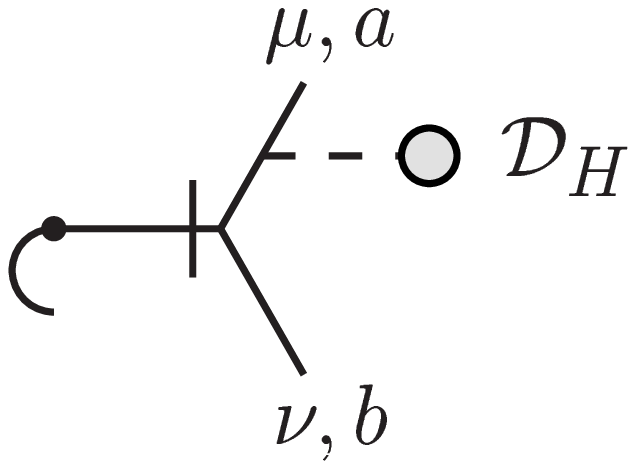}}=4ig^2m^3\f{abi}g^{\mu\nu}\mathcal{D}_H\left(1-\frac{q_b^2-m^2}{(q_a+q_H)^2-m^2}\right), \label{eq:TPV_H_1}\\
    &\adjustbox{valign=c}{\includegraphics[scale=.4]{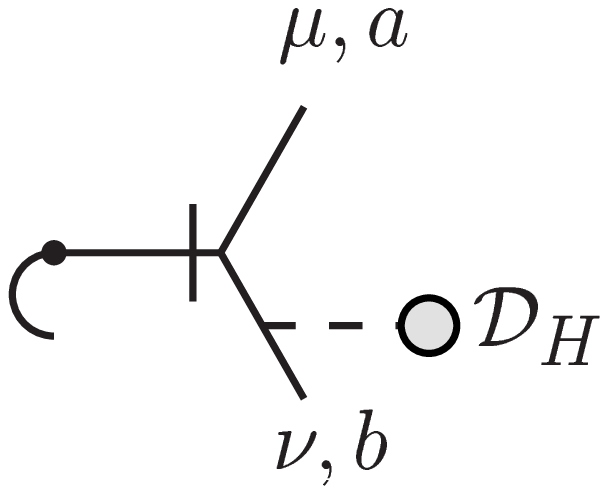}}=4ig^2m^3\f{abi}g^{\mu\nu}\mathcal{D}_H\left(\frac{q_a^2-m^2}{(q_b+q_H)^2-m^2}-1\right) \label{eq:TPV_H_2}
\end{align}
With $q_a$, $q_b$ the momenta of the legs labelled $a$ and $b$ respectively, and $q_H$ the momentum of the Higgs. Moreover, $\mathcal{D}_H$ is the expression resulting from the connected Higgs. When these diagrams are added, the first term of \eqref{eq:TPV_H_1} and the second term of \eqref{eq:TPV_H_2} cancel. It is then relevant what is connected to index $\nu$ and $\mu$ regarding the remaining terms of \eqref{eq:TPV_H_1} and \eqref{eq:TPV_H_2} respectively. For both index $\mu$ and $\nu$ there are three different possibilities, index $\nu$ is considered since index $\mu$ is equivalent due to the symmetry between expressions \eqref{eq:TPV_H_1} and \eqref{eq:TPV_H_2}, minding the relative minus sign. First of all, particle $b$ could be on-shell, i.e. $q_b^2=m^2$, which directly dictates that the remaining term of \eqref{eq:TPV_H_1} is zero. Furthermore, a Yang-Mills vertex could be connected, in which case the cancellation of equation \eqref{eq:TPV_TPV_cancel} or \eqref{eq:TPV_FPV_cancel} transpires. Finally, another Higgs-vector boson coupling can be connected via the $g^{\mu \nu}$ term of the connecting vector bosonic propagator, a cancellation equivalent to that between \eqref{eq:TPV_H_1} and \eqref{eq:TPV_H_2} then occurs, which in turn dictates that the subsequent connection is relevant.\\
Naturally there are more terms, namely the remaining two terms of the TPV and the $q^\mu q^\nu/m^2$ part of the propagators have been neglected. All but one of the terms neglected so far are diagrammatically given by:
\begin{align}
    \includegraphics[scale=.4]{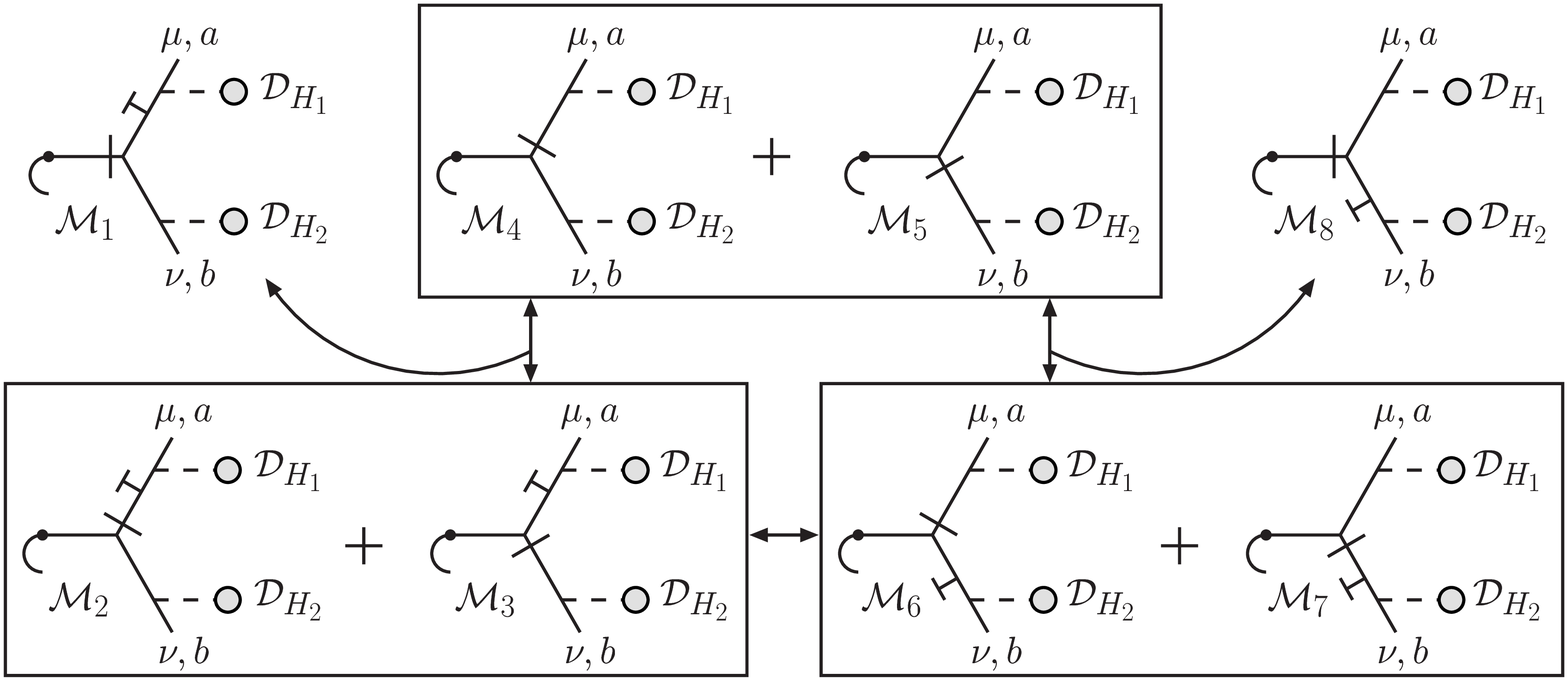}
\end{align}
With $\M_1$ through $\M_8$ denoting the various subdiagrams. The boxed diagrams all vanish completely, with the arrows between the boxes indicating which terms cancel each other. When all terms are added, the only remaining terms arise from $\M_1$ and $\M_8$, these are:
\begin{align}
    &\M_1=ig^3m^3\f{abi}\mathcal{D}_{H_1}\mathcal{D}_{H_2}\frac{1}{(q_a+q_{H_1})^2-m^2}(q_a+q_{H_1})^\mu (q_a+q_{H_1})^\nu,\\
    &\M_8=-ig^3m^3\f{abi}\mathcal{D}_{H_1}\mathcal{D}_{H_2}\frac{1}{(q_b+q_{H_2})^2-m^2}(q_b+q_{H_2})^\mu (q_b+q_{H_2})^\nu
\end{align}
Similar to the cancellation of \eqref{eq:TPV_H_1} and \eqref{eq:TPV_H_2}, there exists a diagram for both $\M_1$ and $\M_8$ that cancels their remaining terms. For $\M_1$ this is:
\begin{align}
    \adjustbox{valign=c}{\includegraphics[scale=.4]{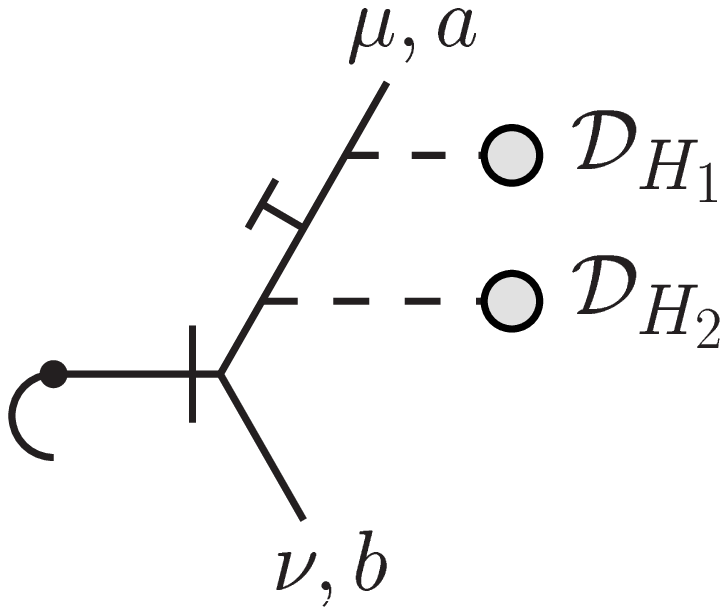}}\begin{array}{l} =ig^2m^3\f{abi}(q_a+q_{H_1})^\mu (q_a+q_{H_1})^\nu \mathcal{D}_{H_1}\mathcal{D}_{H_2}\\ \quad\quad\quad\quad\quad\quad\quad\quad\frac{1}{(q_a+q_{H_1})^2-m^2}\bigg(1-\frac{q_b^2}{(q_a+q_{H_1}+q_{H_2})^2-m^2}\bigg)\end{array} \label{eq:Higgs_counter_term}
\end{align}
In which the first term of expression \eqref{eq:Higgs_counter_term} cancels $\M_1$. Naturally an equivalent diagram exists for $\M_8$. The next cancellation is dependant on what is connected to indices $\mu$ and $\nu$, identical to the cancellations of \eqref{eq:TPV_H_1} and \eqref{eq:TPV_H_2}. The only remaining diagram is a handlebarred TPV in which both connected vector boson propagators carry the $q^\mu q^\nu/m^2$ term, however this term is trivially zero, which can be seen when inspecting the $Y$-function. Finally, no constraints have been placed upon the Higgs momenta, thus the Higgs-Higgs-vector boson coupling is encompassed within this proof. \\
Thus a handlebar on a Yang-Mills vertex is always zero in an $SU(2)$ theory, provided that all vector boson masses are equal. This is of course reminiscent of the Ward-Takahashi identity, differing in two key aspects: this proof pertains to a masive gauge theory, and a handlebarred amplitude is not zero seeing as only Yang-Mills vertices vanish. Naturally the Ward-Takahashi identitiy emerges in the massless limit.\\
Without any further cancellations, maximized power counting dictates that the highest energy dependence of any amplitude is now $E^2$. Consider an amplitude consisting solely of $t$ TPV's and $f$ FPV's, such an amplitude has $t+2f+2$ external particles, and $t+f-1$ propagators. The TPV itself scales with energy, thus, since all $q^\mu q^\nu/m^2$ terms are then zero, power counting gives\footnote{For a fully longitudinally polarized amplitude.}:
\begin{align}
    E^{t+t+2f+2-2t-2f+2}=E^4
\end{align}
However, as given by identity \eqref{eq:Longitudinal_polarization_definition}, the leading part of a longitudinal polarization vector is itself a handlebar, which evaluates to zero. Therefore, the subleading contribution must be taken, which has energy dependence $E^{-1}$, resulting in $E^2$. \\
This counting has, however, neglected any Higgs-vector boson interactions. Fortunately, the total energy contribution of such interactions is zero. When a Higgs, either on- or off-shell, is connected to a vector boson in an arbitrary amplitude the total energy behaviour does not change; an additional vector boson propagator is created by the added Higgs, which provides $E^0$ due to the non-vanishing $q^\mu q^\nu/m^2$ term, while the on- or off-shell Higgs also provides $E^0$. An on-shell Higgs trivially provides an $E^0$ contribution, while an off-shell Higgs connected to two sub-amplitudes has a scaling of $E^2$, which turns to $E^0$ when the Higgs propagator is added. Thus any amplitude has $E^2$ at the highest. Furthermore, a self-interaction appears whenever a Higgs is connected to another Higgs, thereby introducing an additional Higgs-propagator which yields $E^{-2}$. Consequently, diagrams with a single Higgs self-interaction contribute at $E^0$, while any supernumerary Higgs self-interactions have a leading energy contribution lower than $E^0$, and are therefore safe. Thus, with only the aforementioned cancellations, a fully longitudinally polarized amplitude scales as $E^2$ at the highest. Naturally any amplitude at $E^2$ also need to be cancelled, in addition to the $E^0$ terms for amplitudes involving more than four external particles.

\section{Generating functions for off-shell amplitudes}
In order to cancel the $E^2$ terms, the off-shell amplitude is studied at $E^1$ level in which the full propagator of the initial off-shell particle is added. The calculation of all possible processes up to nine external particles via computer algebra show the following pattern for the three vector bosons:
\begin{align}
&\adjustbox{valign=c}{\includegraphics[scale=.5]{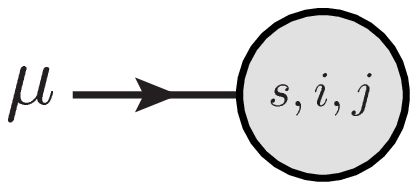}}=\Upsilon^\mu_{W^+},&
&\adjustbox{valign=c}{\includegraphics[scale=.5]{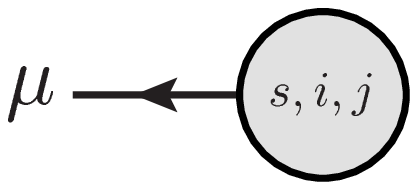}}=\Upsilon^\mu_{W^-},&
&\adjustbox{valign=c}{\includegraphics[scale=.5]{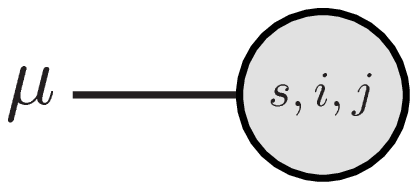}}=\Upsilon^\mu_{Z},&
\end{align}
\vspace{-7pt}
\begin{align}
\Upsilon^\mu_\upsilon = a_\upsilon ^{(s,i,j)}p^\mu +b_\upsilon^{(s,i,j)}m^\mu +c_\upsilon^{(s,i,j)}n^\mu +d_\upsilon^{(s,i,j)} h^\mu \label{eq:upshot_def_VB}
\end{align}
Where $p^\mu$, $m^\mu$, $n^\mu$, and $h^\mu$ are the sums of the $W^+$, $W^-$, $Z$, and $H$ momenta respectively, while $a$, $b$, $c$ and $d$ are integers. Furthermore, $\upsilon$ indicates the off-shell particle and $s$, $i$, and $j$ the number of $W^+$ $W^-$ pairs, $Z$ bosons, and Higgses respectively. An off-shell Higgs provides simply an integer:
\begin{align}
  \adjustbox{valign=c}{\includegraphics[scale=.5]{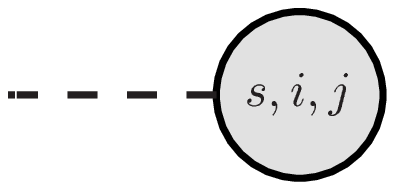}}= \Upsilon_H = e^{(s,i,j)}_H \label{eq:upshot_def_H}
\end{align}
Both expressions \eqref{eq:upshot_def_VB} and \eqref{eq:upshot_def_H} are referred to as the {\em upshot}. The integers of the upshots are then subsumed in thirteen generating functions:
\begin{align}
    &a_{W^+}^{(s,i,j)}\Rightarrow A^{(+)}, & &a_{W^-}^{(s,i,j)}\Rightarrow A^{(-)}, & & a_{Z}^{(s,i,j)}\Rightarrow A^{(0)}, && b_{W^+}^{(s,i,j)}\Rightarrow B^{(+)}, & & b_{W^-}^{(s,i,j)}\Rightarrow B^{(-)},\nonumber \\
    &b_Z^{(s,i,j)}\Rightarrow B^{(0)}, &&c_{W^+}^{(s,i,j)}\Rightarrow C^{(+)}, & &c_{W^-}^{(s,i,j)}\Rightarrow C^{(-)}, &&c_Z^{(s,i,j)}\Rightarrow C^{(0)}, &&d_{W^+}^{(s,i,j)}\Rightarrow D^{(+)}, \nonumber \\
   &d_{W^-}^{(s,i,j)}\Rightarrow D^{(-)}, && d_Z^{(s,i,j)}\Rightarrow D^{(0)}, &&e_H^{(s,i,j)}\Rightarrow E^{(h)} \label{eq:gen_func}
\end{align}
The following convention is used for all generating functions\footnote{The use of $s!^2$ is vital for obtaining sensible generating functions due to the counting of two particles, the $W^+$ and $W^-$ bosons, with a single variable.}:
\begin{align}
    F(t,x,y)=\sum_{s,i,j\geq0}\frac{\zeta^{(s,i,j)}}{s!^2i!j!}t^sx^iy^j
\end{align}
With $\zeta^{(s,i,j)}$ being any of the aforementioned integers of equation \eqref{eq:gen_func}. The conjectured generating functions are constructed by generalizing apparent patterns discerned from the numerically-computed amplitudes. These conjectured generating functions read:
\begin{align}
    &A^{(+)}(t,x,y)=\frac{1+y-x}{\Delta} && A^{(-)}(t,x,y)= 0 && A^{(0)}(t,x,y) = e_t \frac{1}{\Delta} \nonumber\\
    &B^{(+)}(t,x,y)=0 && B^{(-)}(t,x,y)=\frac{1+y+x}{\Delta} && B^{(0)}(t,x,y)= e_t\frac{-1}{\Delta}\nonumber\\
    &C^{(+)}(t,x,y) = e_x \partial_t\frac{t}{\Delta} && C^{(-)}(t,x,y)=e_x \partial_t \frac{-t}{\Delta} && C^{(0)}(t,x,y)= e_x \frac{1+y}{\Delta} \nonumber\\
    &D^{(+)}(t,x,y)= e_y \partial_t \frac{-t}{\Delta} && D^{(-)}(t,x,y)= e_y \partial_t \frac{-t}{\Delta} && D^{(0)}(t,x,y)=e_y \frac{-x}{\Delta} \nonumber\\
    & E^{(h)}(t,x,y)=-1+\sqrt{\Delta} && \Delta = (1+y)^2-x^2-2t &&e_k = \partial^{-1}_k \quad k \in \{x,y,t\} \label{eq:generating_functions_def}
\end{align}
The correctness of these generating functions is proven through the Schwinger-Dyson equation, which, for example, is given by equation \eqref{eq:schwinger-dyson_Z} for the $Z$ boson.
\begin{align}
    \adjustbox{valign=c}{\includegraphics[scale=.5]{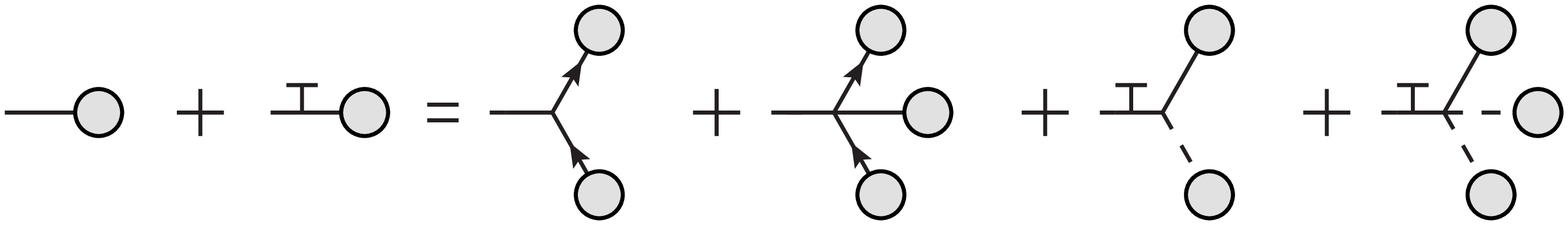}} \label{eq:schwinger-dyson_Z}
\end{align}
The Schwinger-Dyson equations for the $W^+$ and $W^-$ are analogous, differing only in the specific couplings used. If the generating functions are correct for all possible amplitudes, then each connected vector boson provides one momentum, while connected Higgses contribute a scalar. Therefore, every coupling has a total of three momenta; the TPV acquires two momenta from the connected vector bosons in addition to a momentum resulting from the $Y$-function, while an FPV directly obtains three momenta from its connected particles. The Higgs-vector boson couplings have a priori one momentum from its single connected vector boson, but obtain two additional momenta by virtue of the $q^\mu q^\nu/m^2$ term in the propagator, as dictated by power counting. Similarly, an off-shell Higgs is given by the product of two momenta; one from each connected vector boson. Its Schwinger-Dyson equation is:
\begin{align}
    \adjustbox{valign=c}{\includegraphics[scale=.5]{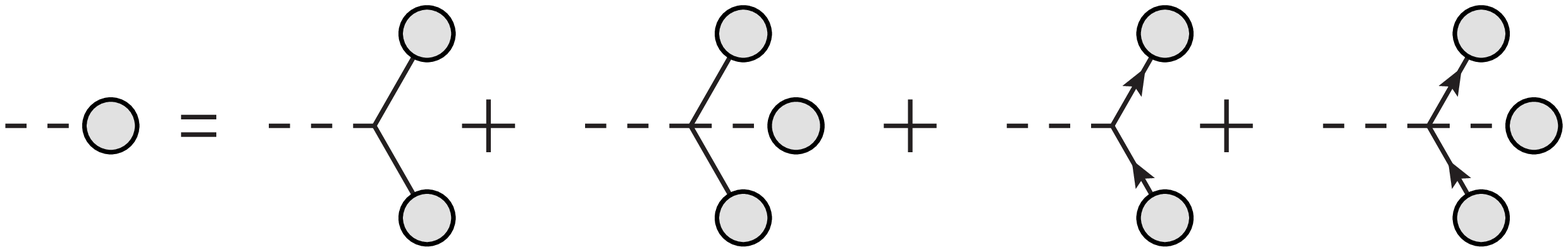}}
\end{align}
The complete Schwinger-Dyson equation is thus given by a product of three momenta for an off-shell vector boson, and two momenta for an off-shell Higgs, both in combination with some to be determined constant given by $\A^{(\iota)}_{ab}$, $\B^{(\iota)}_{ab}$, $\C^{(\iota)}_{ab}$, $\D^{(\iota)}_{ab}$, or $\E^{(h)}_{ab}$, corresponding to $A^{(\iota)}(t,x,y)$, $B^{(\iota)}(t,x,y)$, $C^{(\iota)}(t,x,y)$, $D^{(\iota)}(t,x,y)$, and $E^{(h)}(t,x,y)$ respectively, with $a$, and $b$ specifying the participating momenta, and $\iota$ being either $+$, $-$, or $0$ depending on the investigated generating function. Furthermore, power counting dictates that only the first order in the Taylor expansion of the propagator contributes at $E^2$ level. The Schwinger-Dyson equation of the Higgs thus reads:
\begin{align}
E^{(h)}(t,x,y)=\frac{1}{q^2}\bigg(\E^{(h)}_{pp} p\cdot p + \E^{(h)}_{pm} p\cdot m + \E^{(h)}_{pn} p\cdot n + \E^{(h)}_{ph} p\cdot h + \E^{(h)}_{mm} m\cdot m + \E^{(h)}_{mn} m\cdot n\nonumber\\
+ \E^{(h)}_{mh} m\cdot h + \E^{(h)}_{nn} n\cdot n + \E^{(h)}_{nh} n\cdot h + \E^{(h)}_{hh} h\cdot h\bigg)
\end{align}
The Schwinger-Dyson equations for all other generating functions almost identical, differing only in an additional overall factor of $p^\mu$, $m^\mu$, $n^\mu$, or $h^\mu$ for $A^{(\iota)}(t,x,y)$, $B^{(\iota)}(t,x,y)$, $C^{(\iota)}(t,x,y)$, or $D^{(\iota)}(t,x,y)$ respectively. \\
The complete expression is calculated by summing over all possible distributions of particles in every vertex given by the relevant Schwinger-Dyson equation. The distributions are easily kept track of by employing binomials for three-point vertices and trinomials for four-point vertices, they are respectively defined as:
\begin{align}
    &\binom{a}{b}=\frac{a!}{b!(a-b)!},&&\binom{a}{b,c}=\frac{a!}{b!c!(a-b-c)!}
\end{align}
Restrictions may be imposed upon the bi- and trinomials, depending on the selected product of momenta. For example, the $p \cdot p$ term as given by an off-shell Higgs going to two $Z$'s is: 
\begin{align}
   \adjustbox{valign=c}{\includegraphics[scale=.37]{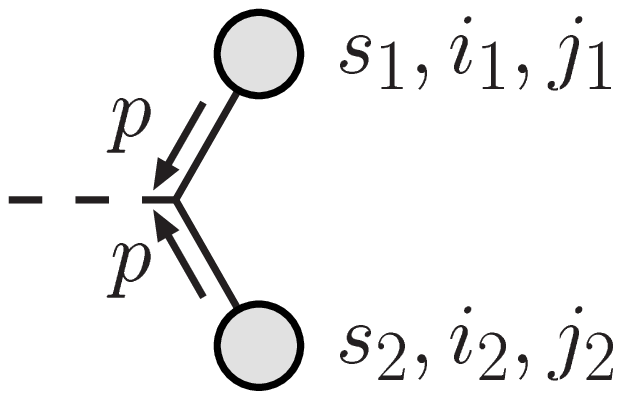}}=\E^{(h)}_{pp_{(ZZ)}}=\frac{1}{2} \sum_{{\tiny  \begin{matrix} s,i,j\geq 0 \\ s_1,i_1,j_1 \geq 0\end{matrix}}} \binom{s-2}{s_1-1}\binom{s}{s_1}\binom{i}{i_1}\binom{j}{j_1} a_Z^{(s_1,i_1,j_1)}a_Z^{(s_2,i_2,j_2)}\Theta \begin{pmatrix} s=s_1+s_2 \\ i=i_1+i_2 \\ j=j_1+j_2 \end{pmatrix}\label{eq:Higgs_pp_binomials}
\end{align}
With $\Theta$ being the logical step function\footnote{$\Theta(P)$ is equal to 1 when $P$ is true, and 0 otherwise}. Naturally, for each particle type there is a binomial, with the $W^+$ and $W^-$ both using the same counting variable $s$, since they are linked through charge conservation. The particles are distributed as given by the binomials, with the selected momenta requiring special attention; here the $p\cdot p$ term dictates that both $Z$'s need to contain at least one $W^+$ each, thus yielding the first binomial. The remaining binomials arise trivially, since no constraints are placed upon the particle type distributions. When equation \eqref{eq:Higgs_pp_binomials} is rewritten in terms of generating functions it reads:
\begin{align}
     \E^{(h)}_{pp_{(ZZ)}}= \frac{1}{2}e_te_t \bigg(\partial_t\frac{a_Z^{(s_1,i_1,j_1)}t^{s_1}x^{i_1}y^{j_1}}{s_1!^2i_1!j_1!}\bigg)\bigg(\partial_t\frac{a_Z^{(s_2,i_2,j_2)}t^{s_2}x^{i_2}y^{j_2}}{s_2!^2i_2!j_2!}\bigg)
   = \frac{1}{2}e_t e_t \bigg(\partial_t A^{(0)}(x,y,t)\bigg)^2
\end{align}
The transition from a bi- or trinomial distribution to a differential equation can be generalised. For distributions in which the relevant variable counts only a single particle type, i.e. the $Z$ and Higgs bosons, the equation becomes:
\begin{align}
    &\sum \binom{l-r}{l_1-r_1,l_2-r_2}\zeta^{(s_1,i_1,j_1)} \zeta^{(s_2,i_2,j_2)} \zeta^{(s_3,i_3,j_3)} \Theta \begin{pmatrix} l=l_1+l_2+l_3 \\ r=r_1+r_2+r_3\end{pmatrix} \nonumber \\
    &=e_l^r \bigg(\partial_l^{r_1} F^{(1)}\bigg)\bigg(\partial_l^{r_2}F^{(2)}\bigg)\bigg(\partial_l^{r_3}F^{(3)}\bigg) \label{eq:upshot_rule_1}
\end{align}
Where $l$ counts the total number of a specified particle type, $l_k$ the number of said particle type in the connected boson labelled $k$, $r$ the number of different momenta that are present in the considered product of momenta, and $r_k$ the number different momenta coming connected boson $k$. Furthermore, $F^{(k)}$ is the relevant generating function related to boson $k$, and $\sum$ indicates a sum over all involved variables, except $r_i$, from 0 to infinity. \\
Since the number of $W^+$'s and $W^-$'s are counted as pairs, via a single variable $s$, the demarcation of four types of vertices is necessitated for writing down the general rules for the resulting differential equations. Those four types are: a vertex that involves no charged particles (e.g. $H\rightarrow ZZ$), a vertex that starts with a $W^+$ or $W^-$ (e.g. $W^+ \rightarrow W^+H$), a vertex with a neutral particle that splits into a $W^+$ $W^-$ pair (e.g. $H\rightarrow W^+W^-$), and finally a vertex that starts with a charged particle and has a charge split (e.g. $W^+\rightarrow W^+W^+W^-$). Let the number of selected $W^+$ momenta be $r$ and $W^-$ momenta $v$. Respectively, the differential equations are then:
\begin{align}
    &\sum \binom{s-r}{s_1-r_1,s_2-r_2}\binom{s-v}{s_1-v_1,s_2-v_2} \zeta^{(s_1,i_1,j_1)} \zeta^{(s_2,i_2,j_2)} \zeta^{(s_3,i_3,j_3)} \Theta \begin{pmatrix}s=s_1+s_2+s_3 \\ r=r_1+r_2+r_3 \\ v= v_1+v_2+v+3 \end{pmatrix}\nonumber\\
    &=e_t^v t^{-v} e_t^r \bigg( \partial_t^{r_1} t^{v_1} \partial_t^{v_1} F^{(1)}\bigg)\bigg( \partial_t^{r_2} t^{v_2} \partial_t^{v_2} F^{(2)}\bigg)\bigg( \partial_t^{r_3} t^{v_3} \partial_t^{v_3} F^{(3)}\bigg), \\
    &\sum\binom{s-r}{s_1+1-r_1,s_2}\binom{s-v}{s_1,s_2+1-v_2} \zeta^{(s_1,i_1,j_1)} \zeta^{(s_2,i_2,j_2)} \zeta^{(s_3,i_3,j_3)}\Theta \begin{pmatrix}s=s_1+s_2+s_3+1\\r=r_1+r_3\\v=v_2+v_3\end{pmatrix}\nonumber\\
    &=e_t^v t^{-v}e_t^r\bigg(\partial_t^{r_1-1}F^{(1)}\bigg)\bigg(t^{v_2-1}\partial_t^{v_2-1}F^{(2)}\bigg)\bigg(\partial^{r_3}t^{v_3}\partial^{v_3}_tF^{(3)}\bigg),\\
    &\sum \binom{s+1-r}{s_1-r_1+1,s_2}\binom{s-v}{s_1,s_2-v_2}\zeta^{(s_1,i_1,j_1)} \zeta^{(s_2,i_2,j_2)} \zeta^{(s_3,i_3,j_3)}\Theta \begin{pmatrix}s=s_1+s_2+s_3\\r=r_1+r_3\\v=v_2+v_3 \end{pmatrix}\nonumber\\
    &=e_t^vt^{-v}e_t^{r-1} \bigg(\partial_t^{r_1-1}F^{(1)}\bigg) \bigg(t^{v_2}\partial_t^{v_2}F^{(2)}\bigg) \bigg( \partial_t^{r_3}t^{v_3} \partial_t^{v_3}F^{(3)}\bigg),\\
    &\sum\binom{s+1-r}{s_1+1-r_1,s_2}\binom{s-v}{s_1,s_2+1-v_2} \zeta^{(s_1,i_1,j_1)} \zeta^{(s_2,i_2,j_2)} \zeta^{(s_3,i_3,j_3)}\Theta \begin{pmatrix}s=s_1+s_2+s_3+1 \\ r=r_1+r_3\\v=v_2\end{pmatrix}\nonumber\\
    &e_t^{v}t^{-v}e_t^r\bigg(\partial_t^{r_1-1}F^{(1)}\bigg) \bigg(t^{v_2-1}\partial_t^{v_2-1}F^{(2)}\bigg) \bigg(\partial_t^{r_3-1}F^{(3)} \bigg) \label{eq:upshot_rule_5}
\end{align}
With of course $e_t^{-1}=\partial_t$ and $\partial_t^{-1}=e_t$. Equations \eqref{eq:upshot_rule_1} through \eqref{eq:upshot_rule_5} completely specify the differential equations resulting from the Schwinger-Dyson equations for given a product of momenta. For example, the complete equation for $\E^{(h)}_{pp}$ is:
\begin{align}
    \E^{(h)}_{pp}=\frac{1}{2}e_t^2\bigg(\partial_t A^{(0)}\bigg)^2\bigg(1+E^{(h)}\bigg) + e_t \bigg(A^{(+)}\bigg)\bigg(\partial_t A^{(-)}\bigg)\bigg(1+E^{(h)}\bigg)=\frac{1}{2}E^{(h)}
\end{align}
Naturally, similar equations exist for all other combinations of momenta. However, the complete set of equations is fairly extensive, containing over 1500 terms. It is for this reason that the entire proof is not provided here, but can be found in\cite{Thesis}. Nevertheless, the complete set of equations prove that $\A^{(\iota)}_{ab}=A^{(\iota)}(t,x,y)$, $\B^{(\iota)}_{ab}=B^{(\iota)}(t,x,y)$, $\C^{(\iota)}_{ab}=C^{(\iota)}(t,x,y)$, $\D^{(\iota)}_{ab}=D^{(\iota)}(t,x,y)$, and $\E^{(h)}_{ab}=E^{(h)}(t,x,y)$, with an exception when $a=b$, in which case an additional factor $\frac{1}{2}$ appears. Thus the conjectured generating functions are correct, and by extension, the upshots of equations \eqref{eq:upshot_def_VB} and \eqref{eq:upshot_def_H} are true for all amplitudes.\\
However, the upshot contains the entire propagator of the initial particle, thus the upshot can never correspond to a completely on-shell amplitude. Therefore $\overline{\Upsilon}_\upsilon^\mu$ is introduced, which indicates the upshot without the initial propagator. A distinction is now made between the first encountered vertices for the upshot of the vector bosons, since a Yang-Mills vertex must be connected to the $-g^{\mu \nu}$ term of the vector bosonic propagator, while all other vertices with $q^\mu q^\nu/m^2$, as dictated by power counting. All diagrams in which the first vertex is a Yang-Mills vertex are subsumed in $\overline{\Upsilon}^\mu_{\upsilon,YM}$, and all other first vertices in $\overline{\Upsilon}^\mu_{\upsilon,V}$, all evaluated at $\mathcal{O}(E^2)$ of course. The upshot for an off-shell Higgs requires no such splitting. This results in:
\begin{align}
    &\overline{\Upsilon}_\upsilon^\mu = \overline{\Upsilon}_{\upsilon,YM}^\mu +\overline{\Upsilon}_{\upsilon,V}^\mu, & &\overline{\Upsilon}_H = -iq^2 \Upsilon_H, \nonumber\\
    &\overline{\Upsilon}_{\upsilon,YM}^\mu = iq^2\Upsilon_\upsilon^\mu-i(q\cdot \Upsilon_\upsilon)q^\mu, & &\overline{\Upsilon}_{\upsilon,V}^\mu = -im^2q^\mu 
\end{align}
Because the upshot contains no poles, the expressions for $\overline{\Upsilon}^\mu_{\upsilon}$ and $\overline{\Upsilon}_H$ remain valid when $q^2\rightarrow m^2$ for all processes. \\
It is already known that $q\cdot \overline{\Upsilon}_{\upsilon,YM} = 0$ from the preceding section, thus when a longitudinal polarization is contracted with $\overline{\Upsilon}_{\upsilon}^\mu$, it must be:
\begin{align}
    &\epsilon_L\cdot \overline{\Upsilon}_{\upsilon} = -t\cdot \overline{\Upsilon}_{\upsilon,YM} + q\cdot \overline{\Upsilon}_{\upsilon,H}-t\cdot \overline{\Upsilon}_{\upsilon,H}=\nonumber \\
    &-im^2(\Upsilon_\upsilon\cdot t) +im^2(q\cdot \Upsilon_\upsilon)-im^2(q\cdot \Upsilon_\upsilon)+im^2(\Upsilon_\upsilon\cdot t) = 0
\end{align}
Additionally, $\overline{\Upsilon}_H\xrightarrow{q^2=m^2}-im^2\Upsilon_H$. Thus all $E^2$ contributions vanish.

\section{Final cancellation mechanisms}
The $E^0$ terms are the only remaining non-safe terms. There are three distinct contributions; the Higgs self-interactions, the higher orders of the Taylor expansion of the vector boson propagator, and either two transverse polarizations or two contributions from the subleading term of the longitudinal polarization.\\
The Higgs self-interactions, in addition to the second order Taylor expansion of the Higgs propagator, provide distinct $M^2$ terms, which, when all summed, yield:\\
\begin{align}
    &\adjustbox{valign=c}{\includegraphics[scale=.5]{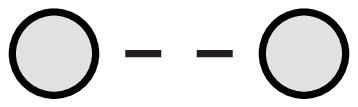}}+
    \adjustbox{valign=c}{\includegraphics[scale=.5]{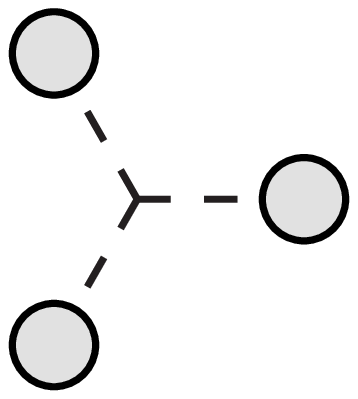}}+
    \adjustbox{valign=c}{\includegraphics[scale=.5]{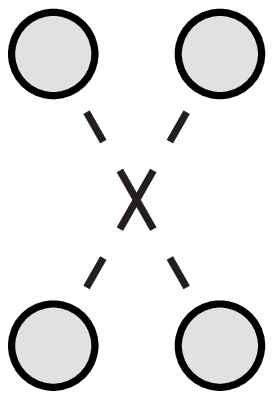}}=
    -\frac{1}{2}{E^{(h)}}^2-\frac{1}{2}{E^{(h)}}^3-\frac{1}{8}{E^{(h)}}^4=-\frac{1}{8}(2y+y^2-x^2-2t)^2
\end{align}
Which shows that only amplitudes consisting of four or fewer particles provide a non-zero contribution, which in turn are the only amplitudes allowed by unitarity to have a non-zero $E^0$ contribution. The $M^2$ terms at $E^0$ are:
\begin{align}
&\mathcal{M}(3H)=-3iM^2      &&\mathcal{M}(4H)=-3iM^2       &&\mathcal{M}(4Z)=-3iM^2\nonumber\\
&\mathcal{M}(2ZH)=iM^2       &&\mathcal{M}(2Z2H)=iM^2       &&\mathcal{M}(W^+W^-2Z)=-iM^2\nonumber\\
&\mathcal{M}(W^+W^-H)=iM^2   &&\mathcal{M}(W^+W^-2H)=iM^2   &&\mathcal{M}(2W^+2W^-)=-2iM^2 
\end{align}
The notation is employed where $\M(sW^+sW^-iZjH)$ indicates an amplitude involving the specified number of particles. These results are easily verified by manual computation of the relevant diagrams. \\
An additional source of $E^0$ terms is when the second order in the Taylor expansion of the vector boson propagator is used for the upshot. Additionally, the $-g^{\mu \nu}$ term in the propagator contracted with $\overline{\Upsilon}^\mu_{\upsilon,V}$ in the first order Taylor expansion also provides a contribution. These terms combine to give:
\begin{align}
    -im^2\frac{g^{\mu\nu}}{q^4}\overline{\Upsilon}_{\upsilon,YM,\nu}+i\frac{q^\mu q^\nu}{q^4}\overline{\Upsilon}_{\upsilon,V,\nu}-i\frac{g^{\mu\nu}}{q^2}\overline{\Upsilon}_{\upsilon,V,\nu} = \frac{m^2}{q^2}\left( \Upsilon_\upsilon^\mu - \frac{(q\cdot \Upsilon_\upsilon)q^\mu}{q^2} +\frac{(q\cdot \Upsilon_\upsilon)q^\mu}{q^2}-\Upsilon_\upsilon^\mu\right) = 0 \label{eq:second_order_taylor}
\end{align}
Naturally, an amplitude contains no Yang-Mills vertices when it consists solely of $Z$ and $H$ bosons, i.e. $\overline{\Upsilon}_{\upsilon,YM,\nu}=0$. However, the upshot will then completely be specified by $C^{(0)}(0,x,y)$ and $D^{(0)}(0,x,y)$, which are then:
\begin{align}
    C^{(0)}(0,x,y)=D^{(0)}(0,x,y)=\frac{1}{2}\ln \left(\frac{1+y+x}{1+y-x}\right)
\end{align}
Thus $\overline{\Upsilon}^\mu_{\upsilon,V}=-im^2\Upsilon_{\upsilon,V}^\mu\rightarrow -im^2q^\mu$, which results in equation \eqref{eq:second_order_taylor} becoming:
\begin{align}
    i\frac{q^\mu q^\nu}{q^4} \overline{\Upsilon}_{\upsilon,V,\nu}-i\frac{g^{\mu\nu}}{q^2}\overline{\Upsilon}_{\upsilon,V,\nu}=\frac{m^2q^2q^\mu}{q^4}-\frac{m^2q^\mu}{q^2}=0
\end{align}
Thus, the only remaining $E^0$ term is given when in the upshot either a single transverse polarization, or a single subleading term of a longitudinal polarization, $-t^\mu$, is present, depending on the amplitude. The resulting upshot will naturally no longer be at $E^1$, but at $E^0$ when a transverse polarization is involved, and $E^{-1}$ for $-t^\mu$. The sole difference between a transverse polarization and the subleading $-t^\mu$ part of a longitudinal polarization is that $\epsilon_T\cdot q = 0$, while $-t\cdot q=-m^2$. However, wherever a $t\cdot q$ exists, there must necessarily be a $q\cdot q$ term with one $q$ being provided by the longitudinal polarization. Since $q\cdot q=m^2$, this term has not been used in any previous cancellations, and thus can be used to cancel $-t\cdot q=-m^2$. Consequently, $t\cdot q$ is effectively zero. Therefore, by proving the nullification of the $E^0$ term of an amplitude with a single transverse polarization, a fully longitudinally polarized amplitude is automatically encompassed within said proof.\\
All two-particle upshots with a transverse polarization are of a comparable form. For example, the upshot of a $W^+$ going to a transversely polarized $W^+$ and longitudinal $Z$ is given by $ \Upsilon_{W^+}^\mu = -q^\mu\pole_n+\epsilon_T^\mu$, in which the convention is adopted where:
\begin{align}
    \pole_i=\frac{\epsilon_T\cdot q_i}{q_T\cdot q_i} \label{eq:pole_def}
\end{align}
When, for now, the $\epsilon_T^\mu$ terms are dropped from the TPV's, and the arrows indicate the transversely polarized particle, the upshots are all given by: \\
\begin{align}
&\adjustbox{valign=c}{\includegraphics[width=2cm,height=2cm]{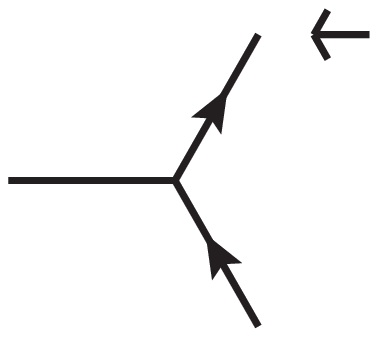}}\Rightarrow -q^\mu\pole_m,   &
&\adjustbox{valign=c}{\includegraphics[width=2cm,height=2cm]{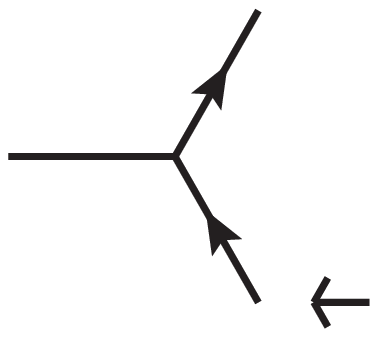}}\Rightarrow  q^\mu\pole_p,   &
&\adjustbox{valign=c}{\includegraphics[width=2cm,height=2cm]{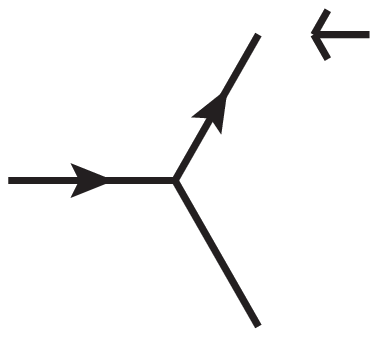}}\Rightarrow  q^\mu\pole_n,   \nonumber\\
&\adjustbox{valign=c}{\includegraphics[width=2cm,height=2cm]{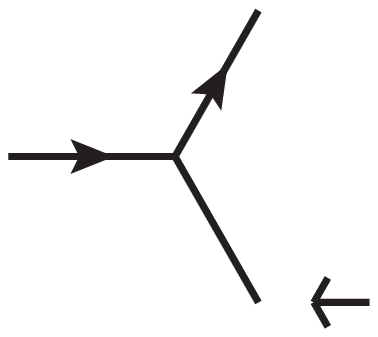}}\Rightarrow -q^\mu\pole_p,   &
&\adjustbox{valign=c}{\includegraphics[width=2cm,height=2cm]{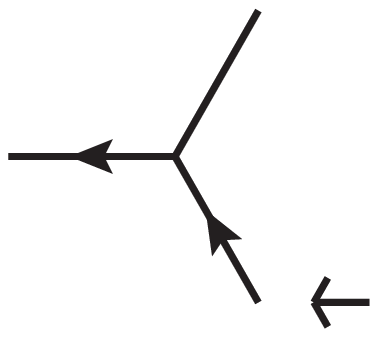}}\Rightarrow -q^\mu\pole_n,   &
&\adjustbox{valign=c}{\includegraphics[width=2cm,height=2cm]{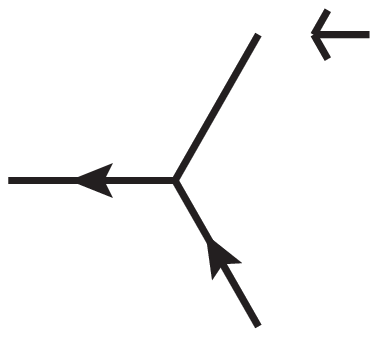}}\Rightarrow  q^\mu\pole_m,   \nonumber\\
&\adjustbox{valign=c}{\includegraphics[width=2cm,height=2cm]{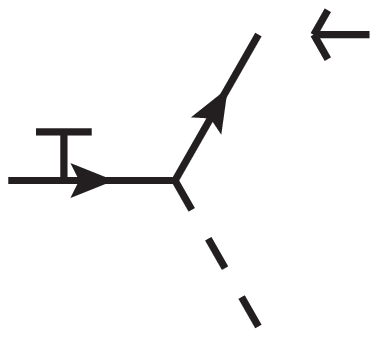}}\Rightarrow -q^\mu\pole_H,   &
&\adjustbox{valign=c}{\includegraphics[width=2cm,height=2cm]{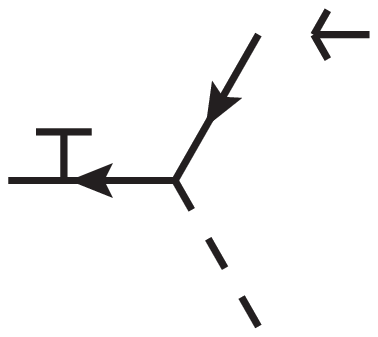}}\Rightarrow -q^\mu\pole_H,   &
&\adjustbox{valign=c}{\includegraphics[width=2cm,height=2cm]{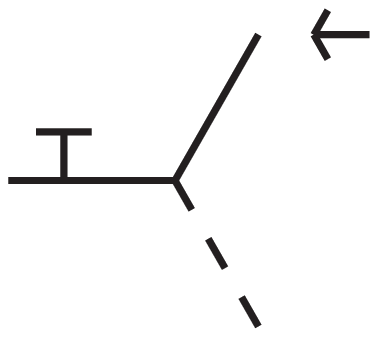}}\Rightarrow -q^\mu\pole_H,   \nonumber\\
&\adjustbox{valign=c}{\includegraphics[width=2cm,height=2cm]{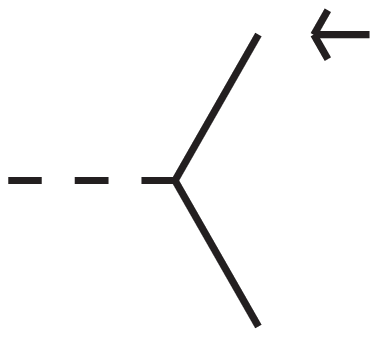}}\Rightarrow       \pole_n,   &
&\adjustbox{valign=c}{\includegraphics[width=2cm,height=2cm]{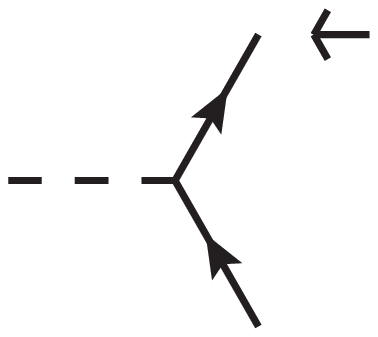}}\Rightarrow       \pole_m,   &
&\adjustbox{valign=c}{\includegraphics[width=2cm,height=2cm]{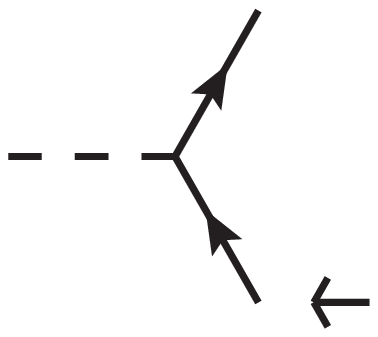}}\Rightarrow       \pole_p   \label{eq:two_pole}
\end{align}
By combining two particles, one of which is transversely polarized, and treating its resulting upshot as an external particle, an $N$-particle amplitude is reduced to an $(N-1)$-particle amplitude of which the upshot is completely known. For example, when a $W^+ W^-$ pair with momenta $p_1^\mu$ and $m_1^\mu$ is contracted into a $H$, its new amplitude is given by $s-1$ $W^+$ $W^-$ pairs and $j+1$ $H$'s, in which the momenta of the new composite Higgs is $p_1^\mu+m_1^\mu$. Since the upshot is known if all external particles are longitudinally polarized, when $\epsilon_T^\mu \rightarrow \epsilon_L^\mu$, the correct upshot must remain, which shall be referred to as the longitudinal limit. By summing all $(N-1)$-particle amplitudes resulting from the relevant two-particle contractions, taking any minus signs resulting from the contraction as dictated by the upshots of equation \eqref{eq:two_pole} into account, and taking the longitudinal limit, the correct upshot of the $N$-particle amplitude is obtained. The constants of all possible $(N-1)$-particle amplitudes are added via the generating functions of \eqref{eq:generating_functions_def}. The addition or removal of a particle via the two-particle contraction is easily accounted for by integration or differentiation respectively. The number of possible contractions is counted by differentiating to the appropriate variable and then multiplying with the same variable:
\begin{align}
    &\frac{\zeta^{(s,i+1,j)}}{s!^2(i+1)!j!}t^sx^{i+1}y^j\Rightarrow \partial_x\frac{\zeta^{(s,i+1,j)}}{s!^2(i+1)!j!}t^sx^{i+1}y^j=\frac{\zeta^{(s,i+1,j)}}{s!^2i!j!}t^sx^iy^j, \\
    &\frac{\zeta^{(s,i-1,j)}}{s!^2(i-1)!j!}t^sx^{i-1}y^j\Rightarrow e_x\frac{\zeta^{(s,i-1,j)}}{s!^2(i-1)!j!}t^sx^{i-1}y^j=\frac{\zeta^{(s,i-1,j)}}{s!^2i!j!}t^sx^iy^j, \\
    &x\partial_x \frac{\zeta^{(s,i,j)}}{s!^2i!j!}t^sx^iy^j = i\frac{\zeta{(s,i,j)}}{s!^2i!j!}t^sx^iy^j
\end{align}
Naturally the addition, removal, or counting of Higgs contractions proceed similarly. The $W^+$ $W^-$ pairs again require special treatment:
\begin{align}
    &\frac{\zeta^{(s-1,i,j)}}{(s-1)!^2i!j!}t^{s-1}x^iy^j\Rightarrow e_t t^{-1} e_t \frac{\zeta^{(s-1,i,j)}}{(s-1)!^2i!j!}t^{s-1}x^iy^j=\frac{\zeta^{(s-1,i,j)}}{s!^2i!j!}t^sx^iy^j
\end{align}
The total proof is again fairly extensive, containing roughly 150 terms, and is thus also not provided here, but is available at\cite{Thesis}. However, a single situation shall be computed for illustration purposes. \\
Consider a process in which a $Z$ goes to a variety of particles specified by $s$, $i$, and $j$, including a single transversely polarized $W^+$. The specific contractions can uniquely be inferred from the particle contents of the $(N-1)$-particle amplitude. For any momentum $p_i^\mu$, excluding the momentum corresponding to the transversely polarized particle, the equations are given by:
\begin{align}
    &Z\rightarrow s W^+,sW^-,iZ,jH & & \\
    &Z\rightarrow (s-1)W^+,(s-1)W^-,(i+1)Z,jH    &&\Rightarrow -t\partial_te_t t^{-1} e_t \partial_x A^{(0)} \pole_m &&=e_t e_t \partial_x \frac{-1}{\Delta}\pole_m\label{eq:upshot_sum_1}\\
    &Z\rightarrow (s-1)W^+,(s-1)W^-,iZ,(j+1)H    &&\Rightarrow -t\partial_t e_t t^{-1} e_t \partial_y A^{(0)} \pole_m&&=e_t e_t \partial_y \frac{-1}{\Delta}\pole_m\\
    &Z\rightarrow sW^+,sW^-,(i-1)Z,jH            &&\Rightarrow x\partial x e_x A^{(0)} \pole_n&&=e_t \frac{x}{\Delta}\pole_n\\
    &Z\rightarrow sW^+,sW^-,iZ,(j-1)H            &&\Rightarrow -y\partial_ye_y A^{(0)} \pole_h&&=e_t \frac{-y}{\Delta}\pole_h\label{eq:upshot_sum_2}
\end{align}
It can easily be seen from identity \eqref{eq:pole_def} that in the longitudinal limit $\pole_i\xrightarrow{\epsilon_T^\mu \rightarrow \epsilon_L^\mu}1$ at leading order. Thus, in this limit, the terms from lines \eqref{eq:upshot_sum_1} through \eqref{eq:upshot_sum_2} can be added, resulting in:
\begin{align}
    e_t e_t \partial_x \frac{-1}{\Delta} + e_t e_t \partial_y \frac{-1}{\Delta}+e_t \frac{x}{\Delta}+e_t \frac{-y}{\Delta} = e_t \frac{1}{\Delta}=A^{(0)}
\end{align}
All individual momenta and their generating functions emerge correctly from this summation, with the exception of the momentum of the transversely polarized particle. Furthermore, since the upshot of the amplitude in the longitudinal limit is known, all other terms must vanish. Consider a single momentum $q_i^\mu$, its form must be a linear combination of terms given by:
\begin{align}
    q_i^\mu \sum_l \frac{\zeta^{(l)} \epsilon_T \cdot q_l}{s_k}
\end{align}
In which $s_k$ is some pole involving $k$ particles. However, in the longitudinal limit all $\zeta^{(l)}$ must vanish except for $k=2$, in which case the correct limit emerges, and since there are no counter terms for $k>2$. Furthermore, any term containing $\epsilon_T^\mu$ cannot contain any poles, again because these poles have to vanish in the longitudinal limit. Moreover, the longitudinal limit dictates the upshot, and therefore the exact form of the term containing $\epsilon_T^\mu$. Thus, the only poles in an amplitude with a single transverse polarization, and therefore a single subleading term of a longitudinal polarization, are those involving two particles. In turn, the exact same cancellation mechanism as for the $E^2$ terms occurs for amplitudes involving more three or more external particles, due to the vanishing of all larger poles. Whereby the nullification of all $E^0$ terms in amplitudes with more than four particles is proven.

\section{Dimensional deformation}
Since all amplitudes, aside those with four or fewer external particles, scale with $E^{-1}$ at the highest, the deformation proof of \cite{abelianhiggs} can be utilized. The proof is included here for completeness. Starting by adding three new dimensions, with the metric being $\hat{g}^{\mu\nu}=\text{diag}[+1,-1,-1,-1,+1,-1,-1]$, and leaving the newly added dimensions empty in the existing momenta and $t$-vectors:
\begin{align}
    q_i^\mu = (q^0_i,q^1_i,q^2_i,q^3_i) \rightarrow (q^0_i,q^1_i,q^2_i,q^3_i,0,0,0) \quad \quad
    t_i^\mu = (t^0_i,t^1_i,t^2_i,t^3_i) \rightarrow (t^0_i,t^1_i,t^2_i,t^3_i,0,0,0)
\end{align}
Furthermore, a new set of vectors,  $r_i^\mu$, are introduced:
\begin{align}
    r_i^\mu = (0,0,0,0,r^4_i,r^4_i,r^5_i)
\end{align}
With the following two conditions imposed upon the set of $r_i^\mu$:
\begin{align}
    r_i\cdot r_i = 0 \quad\forall_i \quad \quad\text{and}\quad \quad \sum_i r_i^\mu = 0
\end{align}
The momenta are then deformed as follows:
\begin{align}
    q^\mu_i\rightarrow \hat{q}^\mu_i =q_i^\mu + z^{\frac{1}{2}}r^\mu_i
\end{align}
With $z$ a free complex parameter. The conditions imposed upon the set of $r_i^\mu$ vectors and the specific deformation ensure that momentum conservation is still valid and that all external particles are still properly on-shell. \\
An $N$-particle amplitude can always be divided into two parts, containing $N_s$ and $N-N_s$ particles, connected via a propagator whose denominator is:
\begin{align}
    \hat{\Delta}_s=(\hat{q}_s^2-m^2)=(q^2_s +z r_s^2 -m^2)=\Delta_s+zr_s^2
\end{align}
Here $m$ can either be the vector boson or Higgs mass, depending on the connecting particle. Naturally, when $z=-\frac{\Delta_s}{r_s^2}$ a pole emerges, its corresponding residue is then denoted by $R(z_s)$. Furthermore, since all amplitudes except those with four or fewer external particles scale with $E^{-1}$ at the highest, the following relations holds:
\begin{align}
    \M_\infty \equiv \lim_{z\rightarrow \infty} \M(z) =\left\{\begin{array}{ll} \text{constant} & n=4 \\ 0 & n\geq4\end{array} \right.
\end{align}
The following contour integral can now be performed:
\begin{align}
    \M(0) = \frac{1}{2\pi i} \oint_{z\sim 0}\frac{\M(z)}{z}dz = \M_\infty -\sum_s\frac{1}{2\pi i}\oint_{z\sim z_s} \frac{R(z_s)}{z\hat{\Delta}_s}dz= \M_\infty +\sum_s \frac{R_s(z_s)}{\Delta_s}
\end{align}
When $z=z_s$ the deformed momentum $\hat{q}_s^\mu$ is on-shell. Combined with the fact that the numerator of the vector bosonic propagator is a sum of the various polarization vectors, which now number six instead of three due to the added dimensions:
\begin{align}
    -\hat{g}^{\mu\nu}-\frac{\hat{q}_s^\mu \hat{q}_s^\nu}{m^2}=\sum^6_{\lambda=1} \hat{\epsilon}^\mu_{s,\lambda}\hat{\epsilon}^\nu_{s,\lambda}
\end{align}
The residue $R(z_s)$ is thus given by $R(z_s)=A_{N_s+1}B_{N-N_s+1}$, in which $A_{N_s+1}$ and $B_{N-N_s+1}$ are on-shell amplitudes with $N_s+1$ and $N-N_s+1$ particles respectively. Thus, assuming that both $A_{N_s+1}$ and $B_{N-N_s+1}$ obey unitarity, the unitarity of the entire amplitude follows inductively:
\begin{align}
    \frac{R(z_s)}{\Delta_s}\rightarrow E^{-2} E^{4-N_s-1} E^{4-N+N_s-1}=E^{4-N}
\end{align}
Thus the unitarity of all possible processes in a spontaneously broken $SU(2)$ theory is proven.

\section{Conclusion}
Within this paper the proof for the unitarity of a massive $SU(2)$ theory at tree level is laid out in the unitary gauge, i.e solely using physical degrees of freedom. While the unitarity of a spontaneously broken $SU(2)$ theory is not particularly surprising, it is the cancellation mechanisms and results of the computation of the Feynman diagrams that are of interest. It is shown that the Ward-Takahashi identity can be extended to include a massive $SU(2)$ theory, albeit in a modified form; only the subset of diagrams of which the first vertex is a Yang-Mills vertex vanish when contracted with a handlebar. Furthermore, thirteen generating functions have been computed that completely specify the linear combination of momenta that result from any off-shell amplitude at $\mathcal{O}(E^1)$ for a spontaneously broken $SU(2)$ theory. The correctness of these generating functions has been proven, thereby showing that no amplitude at $\mathcal{O}(E^2)$ contains any poles, which by extension proves the nullification of all $E^2$ terms. These results greatly resemble previously obtained results for a massive $U(1)$ theory, implying a similar structure for other $SU(N) \quad (N\geq 3)$ theories. The proof is completed with the cancellation of the $E^0$ terms, which is shown to occur via three separate mechanisms. First, the contributions provided by the Higgs self-interactions all vanish, aside from those involving four or fewer external particles. Second, the contraction of $\overline{\Upsilon}_{\upsilon,V,\mu}$ with the $g^{\mu\nu}$ term of the vector boson propagator cancels against the second order term of the Taylor-expanded propagator. Third, any $N$-particle amplitude involving a single transverse polarization at $\mathcal{O}(E^0)$ is shown to be reducible to a linear combination of fully longitudinally polarized $N-1$-particle amplitudes, thereby showing the cancellation of the $E^0$ terms in the exact same manner as the $E^2$ terms. By proving no terms exist above $\mathcal{O}(E^{-1})$, aside from those involving four or fewer external particles, the deformation proof could be used to inductively prove the unitarity of a spontaneously broken $SU(2)$ theory.

\bibliographystyle{unsrt}
\bibliography{main}

\end{document}